\documentclass[reqno]{amsart}
\usepackage{a4wide}

\numberwithin{equation}{section}

\newcommand\Pone{\ensuremath{\mathbf{P}_1}}
\newcommand\pvplus{\ensuremath{\mathbf{P}(\mathbb{V}^+)}}
\newcommand\pv{\ensuremath{\mathbf{P}(V)}}

\theoremstyle{plain}
\newtheorem{theorem}{Theorem}[section]

\theoremstyle{remark}

\theoremstyle{definition}
\newtheorem{definition}{Definition}[section]

\newtheorem*{ack}{Acknowledgements}

\begin{document}

\title{Paraconformal structures and integrable systems}
\author{James D.E. Grant}
\date{15 June 1999}
\thanks{Submitted to \textit{Nonlinearity}}
\thanks{solv-int/9906008}
\address{Department of Mathematics\\ 
University of Hull\\ Hull HU6 7RX\\ U.K.}
\email{j.d.grant@maths.hull.ac.uk}

\begin{abstract}
We consider some natural connections which arise between
right-flat $(p, q)$ paraconformal structures and integrable
systems. We find that such systems may be formulated in Lax form,
with a \lq\lq Lax $p$-tuple\rq\rq\ of linear differential
operators, depending a spectral parameter
which lives in $(q-1)$-dimensional
complex projective space. Generally, the differential operators
contain partial derivatives with respect
to the spectral parameter.
\end{abstract}

\maketitle

\tableofcontents

\section{Introduction}
\label{sec:intro}

It has long been known that in four-dimensional Riemannian
geometry there is a connection between conformal structures with
anti-self-dual Weyl tensor, and $3$-dimensional complex manifolds:
\begin{theorem}[AHS]
\label{thm:ahs}
If a manifold $M$ admits a conformal structure
with $W^+ = 0$, then the projective spin-bundle
{\pvplus} is a complex $3$-manifold. Conversely,
given a complex $3$ manifold $Z$ with a real structure
(i.e. anti-holomorphic involution) $\sigma$ and a $4$ parameter
family of embedded rational curves with normal bundle
$N \cong H \oplus H$, on which $\sigma$ acts as the anti-podal
map, then the space of real rational curves admits an
anti-self-dual conformal structure. All anti-self-dual
conformal structures arise in this way.
\end{theorem}
Moreover, if one considers anti-self-dual Yang-Mills fields on an
anti-self-dual background, then solutions may be constructed in terms of
holomorphic vector bundles on this complex $3$-manifold. The existence
of such a complex $3$-manifold is looked on as being central to the
notion of integrability of the anti-self-duality equations \cite{W1,W2,MW}.

If one tries to generalise notions of anti-self-duality
to higher dimensional Riemannian
manifolds, there are several (inequivalent)
paths one may choose to follow.

A common choice is to investigate Riemannian manifolds
in higher dimension with reduced holonomy group \cite{S},
and gauge fields related to these structures \cite{CDFN}.
In the case of (irreducible, non-symmetric)
Riemannian manifolds, one may then invoke
Berger's classification of the possible holonomy groups
of the Levi-Civita connection.
Apart from the generic case of $\mathrm{SO}(n)$ holonomy group,
the holonomy groups allowed by Berger's classification
correspond to K{\"a}hler, quaternionic-K{\"a}hler, 
Ricci-flat K{\"a}hler and
hyper-K{\"a}hler manifolds, along with two exceptional
possibilities of Ricci-flat metrics with
holonomy group $\mathrm{G}_2$ and
$\mathrm{Spin}_7$ in dimensions $7$ and $8$.
This may not be the
most natural approach if one's interest is in
integrable systems, however, since, of these
possibilities, the only systems which appear to be
integrable are those which govern
K{\"a}hler, quaternionic-K{\"a}hler
and hyper-K{\"a}hler structures. The equations for
Ricci-flat K{\"a}hler metrics are not integrable in
dimensions greater than $4$. Little is
known concerning the integrability of the $\mathrm{G}_2$ and
$\mathrm{Spin}_7$ holonomy equations, although they contain
as special cases the (non-integrable) equations for $6$-dimensional and
$8$-dimensional Ricci-flat K{\"a}hler structures respectively.
Therefore, it is extremely unlikely that these two systems
are integrable.

An alternative path is not to start with the geometrical
condition of anti-self-duality of the $4$-dimensional metric,
but to simply consider systems in higher dimensions where there
is a suitable generalisation of the complex $3$-manifold which
appears in four dimensions. Generically, we will denote such a
complex manifold by $Z$, and the idea is that
one reconstructs the geometrical manifold, denoted $M$,
as a parameter space of particular sub-manifolds
of $Z$. Natural geometrical structures then arise on the manifold
$M$ as a result of the integrability of the complex structure on
$Z$. Any additional holomorphic structures that exist on $Z$ then
lead to more specialised geometrical structures on $M$. The
complex manifold $Z$ will be referred to as a twistor space,
and the essence of the work of Ward and others
is that it is the existence of a twistor construction for a problem
which should be interpreted as a sign of its integrability \cite{W1,W2,MW}.
Substantial evidence for this claim comes from the fact that many
standard integrable systems in $2$ and $3$ dimensions may be
constructed as symmetry reductions of the equations for anti-self-dual
Yang-Mills fields and anti-self-dual conformal structures in $4$
dimensions \cite{W2,MW}. In this approach it is the
existence of the complex manifold $Z$ that is central, the equations on
the space-time then being simply a manifestation of the complex structure
on $Z$.

We aim here to study (local) properties of structures
for which there is a complex manifold construction, the right-flat
paraconformal structures, and to see what new features of integrable
systems these structures suggest. In particular,
we begin by reviewing, in explicitly local terms, the construction
implicit in Theorem~\ref{thm:ahs} above. What we find is that even in
this simplest situation, there are features of the equations which arise
which are unusual from the point of view of integrable systems.
In the generic
case of an anti-self-dual
manifold, the spin-bundle
does not fibre over \Pone, the
complex projective line. In integrable systems terms this means that the
operators in our Lax Pair contain partial derivatives with respect to
the spectral parameter. Therefore the spectral
parameter itself is very much
part of the geometrical problem,
a property which is unusual (but not unknown) in conventional integrable
systems theory.

We then study, in a similar fashion, $(p, q)$ right-flat
paraconformal structures
in dimension $n = p q$. In this case we find that, as opposed to a Lax
pair of operators depending on a spectral parameter $\lambda \in \Pone$,
the right-flat condition on a paraconformal structure is determined by
a Lax $p$-tuple of differential operators depending on a spectral
parameter taking values in the higher-dimensional projective space
$\mathbf{P}_{q-1}$. As in the case of anti-self-dual structures in four
dimensions, these differential operators generally contain derivatives
in the spectral parameter, corresponding to the fact that the complex
manifold $Z$ does not fibre holomorphically over $\mathbf{P}_{q-1}$.

The moral of our story is that if one takes the ideas of Ward and others
seriously, that it is the connection with complex manifold theory which
is central to integrable system theory, then one must substantially
generalise what one considers to be an integrable system.

\section{Anti-self-dual Conformal Structures}
\label{sec:asd}

We begin by reviewing, in local terms, the construction
implicit in Theorem~\ref{thm:ahs} above.
Consider an oriented Riemannian
four-manifold $M$.
We may then define a canonical
almost-complex-structure on the
projective spin-bundle. First
we use the Levi-Civita connection
to split the tangent bundle of {\pvplus} as
the direct sum of a vertical part along
the fibres, $V(\pvplus)$, and the horizontal
part, $H(\pvplus)$, which is the pull-back of the
tangent bundle of $M$, $p^* TM$. The vertical fibres are
complex projective lines, and
so inherit a natural almost complex structure.
In the horizontal direction, a non-zero spinor
$\pi \in (\mathbb{V}^+)_x$ identifies $T_x M$
by Clifford multiplication
with the two-dimensional
complex vector space $(\mathbb{V}^-)_x$. At the
points of {\pvplus} corresponding to $\pi$ we
put this almost complex structure on $H_x(\pvplus)$.
It follows that this
almost complex structure is integrable if and
only if the Weyl tensor of the
conformal structure is anti-self-dual \cite{AHS} .

To cast this in more explicit
terms, fix a Riemannian metric,
$\mathbf{g}$, in the conformal
structure. If we complexify the
tangent space, and extend
the metric by complex linearity to a
complex metric (again denoted
$\mathbf{g}$) on $TM \otimes \mathbb{C}$,
then, locally, we may
introduce a null basis
$\{ \boldsymbol{\epsilon}^i | i = 1, \dots 4 \}$
for $T^* M \otimes \mathbb{C}$ in which the metric
may be written
\begin{equation}
\mathbf{g} =
\boldsymbol{\epsilon}^1 \otimes \boldsymbol{\epsilon}^2 +
\boldsymbol{\epsilon}^2 \otimes \boldsymbol{\epsilon}^1 +
\boldsymbol{\epsilon}^3 \otimes \boldsymbol{\epsilon}^4 +
\boldsymbol{\epsilon}^4 \otimes \boldsymbol{\epsilon}^3.
\label{gee}
\end{equation}

We can then define the Levi-Civita connection,
$\boldsymbol{\Gamma}$, of
the tetrad by the equation
\[
d \boldsymbol{\epsilon}^i +
\sum_{j=1}^4
\boldsymbol{\Gamma}^i{}_j \wedge
\boldsymbol{\epsilon}^j = 0,
\qquad i = 1, \dots , 4.
\]
If we adopt an affine complex
coordinate, $\lambda$, on the fibre $({\pvplus})_x \cong
\Pone$, then we define an almost complex structure
on {\pvplus} by defining the distribution
$\Lambda \subset T^* (\pvplus)$ spanned by the $1$-forms
\begin{align*}
\boldsymbol{\sigma}_1 &=
\boldsymbol{\epsilon}^3 + \lambda
\boldsymbol{\epsilon}^1,
\\
\boldsymbol{\sigma}_2 &=
\boldsymbol{\epsilon}^2 - \lambda
\boldsymbol{\epsilon}^4,
\\
\boldsymbol{\sigma}_3 &=
d\lambda+ \Gamma_{14} + \lambda \left( \Gamma_{12} -
\Gamma_{34} \right) + \lambda^2 \Gamma_{23}.
\end{align*}
This almost complex structure on {\pvplus} is integrable
if and only if the distribution $\Lambda$ is involutive,
i.e. $d \Lambda \subset \Lambda^1 \wedge \Lambda$. It is
straightforward to show that this is the case if and only if the
Weyl tensor of the metric $\mathbf{g}$ defined above is anti-self-dual.
It also follows straightforwardly that this construction is unaffected
by conformal changes of metric, and so depends only on the conformal
equivalence class of the metric \cite{AHS}.

The connection with integrable
systems comes from taking a dual
formulation of this result.
The anti-holomorphic tangent space of
{\pvplus} is spanned by the vector fields
\begin{align*}
\mathbf{v}_1 &= \frac{1}{1 + \lambda{\overline \lambda}}
\left[
\mathbf{e}_4 + A_4 \frac{\partial}{\partial \lambda}
+ {\overline A}_4
\frac{\partial}{\partial {\overline \lambda}} +
\lambda \left(
\mathbf{e}_2 + A_2 \frac{\partial}{\partial \lambda}
+ {\overline A}_2
\frac{\partial}{\partial {\overline \lambda}}
\right) \right],
\\
\mathbf{v}_2 &= \frac{1}{1 + \lambda {\overline \lambda}}
\left[
\mathbf{e}_1 + A_1 \frac{\partial}{\partial \lambda}
+ {\overline A}_1
\frac{\partial}{\partial {\overline \lambda}} -
\lambda \left(
\mathbf{e}_3 +
A_3 \frac{\partial}{\partial \lambda} +
{\overline A}_3
\frac{\partial}{\partial {\overline \lambda}}
\right) \right],
\\
\mathbf{v}_3 &=
\frac{\partial}{\partial {\overline \lambda}},
\end{align*}
where $\mathbf{e}_i$ are vector
fields on $M$ dual to the $1$-forms
$\boldsymbol{\epsilon}^i$
\[
< \boldsymbol{\epsilon}^i, \mathbf{e}_j >
= \delta^i_j,
\]
and
\begin{align*}
\mathbf{A} &=
- \Gamma_{14} - \lambda \left( \Gamma_{12} -
\Gamma_{34} \right) - \lambda^2 \Gamma_{23},
\\
{\overline{\mathbf{A}}} &=
- \Gamma_{23} + \lambda \left( \Gamma_{12} -
\Gamma_{34} \right) - \lambda^2 \Gamma_{14}.
\end{align*}
The complex structure defined by
these vectors is integrable if they are
closed under Lie Brackets. We now
note that the complex structure
defined by these vector fields is the same
as defined by the following basis:
\begin{align*}
\mathbf{L}_1 &= D_4 + \lambda D_2,
\\
\mathbf{L}_2 &= D_1 - \lambda D_3,
\\
\mathbf{v} &=
\frac{\partial}{\partial {\overline \lambda}},
\end{align*}
where we have defined the vector fields
\begin{equation}
D_i = \mathbf{e}_i +
A_i \frac{\partial}{\partial \lambda}.
\label{diffops}
\end{equation}

The only non-trivial part of
the integrability of the complex structure
we have defined is that the
Lie-Bracket of $\mathbf{L}_1$ and
$\mathbf{L}_2$ must lie in
$T^{(0, 1)}$. Therefore for integrability we
require the existence of functions
$\alpha (x: \lambda), \beta (x: \lambda)$ with
the property that
\begin{equation}
\left[ D_4 + \lambda D_2,
D_1 - \lambda D_3 \right] =
\alpha \left( D_4 + \lambda D_2 \right) +
\beta \left( D_1 - \lambda D_3 \right).
\label{alg}
\end{equation}
A power counting argument implies that
the functions $\alpha$ and $\beta$ are
quadratic polynomials in the variable $\lambda$.
If this condition is satisfied,
then the projective spin-bundle is a
complex $3$ manifold, and so the
conformal structure must be
anti-self-dual. Conversely, if
the conformal structure is
anti-self-dual, then the projective
spin-bundle is a complex $3$
manifold and so, locally, we may
choose bases where the above equations
are satisfied. We therefore have:
\begin{theorem}
Given an anti-self-dual conformal
structure and any representative
metric in the conformal class written
in the form \eqref{gee}, then there
exists a $1$-form $\mathbf{A}$, which
is a quadratic function of an arbitrary
\Pone-valued parameter $\lambda$ and two quadratic
functions of $\lambda$, $\alpha$ and $\beta$ which
obey Eqs.~\eqref{alg}, where the differential
operators $D_i$ are as in Eq.~\eqref{diffops}.
\end{theorem}

It is possible to decompose Eqs.~\eqref{alg} into
components in the tangent
space of the manifold $M$ and components in
the vertical direction
$\partial / \partial \lambda$.
The components in $TM$ tell use that
the functions $\alpha, \beta$ and the
components of the form $\mathbf{A}$ correspond to parts of the Levi-Civita
connection. The parts of
the Levi-Civita connection they define are
precisely the parts required to
construct the self-dual part of the
Weyl tensor, ${}^+ W$. The vertical component
of Eqs.~\eqref{alg} then tell
us that the $5$ individual components
of ${}^+ W$ vanish identically, so the Weyl tensor
is anti-self-dual.

Equation~\eqref{alg} tells us that the operators $\mathbf{L}_1$ and
$\mathbf{L}_2$ constitute a Lax pair for the problem, and therefore that
the system is integrable. However, these operators contain derivatives
with respect to the spectral parameter $\lambda$, a feature which does not
usually occur in standard integrable systems theory. The origin of these
derivative terms lies in the nature of the complex manifold \pvplus.
Eqs.~\eqref{alg} are the integrability condition which ensures
the existence of three linearly independent
solutions of the over-determined set of
equations for a function $f(x: \lambda)$
\[
\mathbf{L}_1 f = \mathbf{L}_2 f = 0.
\]
Solutions of these equations correspond to meromorphic
functions on {\pvplus}. The fact that
$\lambda$ itself is not a solution of these
equation is a consequence of the fact that generally {\pvplus} does not
fibre over \Pone (equivalently $\lambda$ is not a
meromorphic function on {\pvplus}). In the case of hyper-K{\"a}hler
or hyper-complex structures, where the
spin-bundle does fibre over \Pone, the $\lambda$ derivatives are not present
in the Lax pair \cite{MN,GS1}. In these cases,
one can reconstruct the transition
functions of the bundle from the solutions
of the above equations \cite{NPT}.

Although Eq.~\eqref{alg} describes the most general anti-self-dual
conformal structures locally, there are various special cases of these
equations:
\begin{itemize}

\item[$\bullet$] Letting
$A_i= \lambda \phi_i$, we recover the class of
Hermitian anti-self-dual spaces,
which are conformal to scalar-flat
${\overline\partial}$-K{\"a}hler metrics \cite{Pr};

\item[$\bullet$] Letting
$A_4 + \lambda A_2 = A_1 - \lambda A_3 = 0$ defines
hyper-complex structures in four dimensions \cite{GS1};

\item[$\bullet$] Letting $A_i = \lambda \phi_i$,
and assuming the vector fields
$\mathbf{e}_i$ are divergence free
with respect to some volume element
defines a scalar-flat K{\"a}hler metric
up to a known conformal factor (this is an
extension of a result of Park \cite{P});

\item[$\bullet$] Letting
$A_4 + \lambda A_2 = A_1 - \lambda A_3 = 0$ and
assuming the vector fields $\mathbf{e}_i$
are divergence free with respect to some volume
element defines a hyper-K{\"a}hler
metric up to a known conformal
factor \cite{MN}.
\end{itemize}

Similar results
hold for complex anti-self-dual
conformal structures and real
conformal structures of signature
$(-,-,+,+)$ with suitable
generalisations and modifications of the
reality conditions.

\section{Paraconformal Structures}
\label{sec:para}

Anti-self-dual conformal structures
in four dimensions are a special
case of a more general type of
structure, a right-flat paraconformal structure
\cite{BE}. Recall that, for integers $p, q \ge 2$,
a $(p, q)$ paraconformal structure consists of a
complex manifold $M$ of complex dimension $n=pq$ and an
isomorphism $\alpha$ between the (holomorphic) tangent
bundle of $M$ and the tensor product of
a rank $p$ complex vector bundle $U$ with a rank
$q$ vector bundle $V$
\begin{equation}
\alpha: TM \to U \otimes V.
\label{paraisom}
\end{equation}
Given such an isomorphism, we may introduce an isomorphism
\begin{equation}
\Lambda^p U \cong \Lambda^q V
\label{isom}
\end{equation}
between the highest exterior powers of these bundles.

Given connections, both denoted $\nabla$, on the bundles
$U$ and $V$, we may define a unique induced connection
on $TM$, again denoted $\nabla$, by demanding that
covariant differentiation commutes with the isomorphism
$\alpha$. This affine connection naturally has torsion
$\mathbf{T}$ defined by
\[
\nabla_{\mathbf{X}} \mathbf{Y} - \nabla_{\mathbf{Y}} \mathbf{X} -
{[ \mathbf{X}, \mathbf{Y} ]}
= \mathbf{T}( \mathbf{X}, \mathbf{Y}),
\qquad
\forall \mathbf{X}, \mathbf{Y} \in \Gamma ( TM )
\]
and curvature tensor $\mathbf{R}$ given by
\[
\left(
\left[ \nabla_{\mathbf{X}}, \nabla_{\mathbf{Y}} \right] -
\nabla_{[ \mathbf{X}, \mathbf{Y} ]}
\right) \mathbf{V} = \mathbf{R}( \mathbf{X}, \mathbf{Y} ) \mathbf{V},
\qquad
\forall \mathbf{X}, \mathbf{Y}, \mathbf{V} \in \Gamma ( TM ).
\]
A \emph{scale} for a paraconformal structure consists of a non-vanishing
section, $\boldsymbol{\epsilon}$, of the bundle $\Lambda^p U$. The
isomorphism~\eqref{isom} then implies the existence of a non-vanishing
section, $\boldsymbol{\epsilon^{\prime}}$, of the bundle $\Lambda^q V$. It
can be shown \cite{BE} that, given the sections $\boldsymbol{\epsilon}$
and $\boldsymbol{\epsilon^{\prime}}$, there exist unique connections on
$U$ and $V$ (and therefore on $TM$) with the property that the torsion
is trace-free, and which annihilate the forms $\boldsymbol{\epsilon}$
and $\boldsymbol{\epsilon^{\prime}}$:
\[
\nabla \boldsymbol{\epsilon} = \nabla \boldsymbol{\epsilon^{\prime}} =
0.
\]
We shall generally assume the existence of a scale, and work with
the unique connections which preserve it.

\subsection{Algebraic Decomposition of the Torsion and Curvature}
\label{sec:alg}
If we consider the space of $2$-forms on $M$, we have
\[
\wedge^2 (T^*M) \cong \wedge^2 (U^* \otimes V^*) \cong
\left( \wedge^2 (U^*) \otimes S^2 (V^*) \right)
\oplus
\left( S^2 (U^*) \otimes \wedge^2 (V^*) \right),
\]
where $\wedge$ and $S$ denote skew-symmetric and symmetric powers of the
relevant bundles, respectively. Viewing the torsion as a $TM$ valued
$2$-form on $M$, it decomposes into two parts
\[
\mathbf{T} = \mathbf{T}^+ \oplus \mathbf{T}^-,
\]
where
\[
\mathbf{T}^+ \in TM \otimes \left( \wedge^2 (U^*) \otimes S^2 (V^*) \right),
\qquad
\mathbf{T}^- \in TM \otimes \left( S^2 (U^*) \otimes \wedge^2 (V^*) \right).
\]
One can show that the trace-free parts of these parts of the
torsion are independent of the connection chosen on the
vector bundles $U$ and $V$ \cite{BE}.

There is a similar decomposition of the curvature tensor $\mathbf{R}$.
Given the direct product nature of $TM$, the curvature decomposes as a
direct sum
\[
\mathbf{R} = \mathbf{R}^- \otimes \mathrm{Id}_V +
\mathrm{Id}_U \otimes \mathbf{R}^+,
\]
where $\mathbf{R}^-$ and $\mathbf{R}^+$ denote the curvatures of the
connections $\nabla$ on $U$ and $V$ respectively. $\mathbf{R}^+$ may be
viewed as a section of $\Lambda^2 (M) \otimes \mathrm{End} (V)$, and
\begin{align*}
\Lambda^2 (M) \otimes \mathrm{End} (V) &\cong
\left(
\left( \Lambda^2 U^* \otimes S^2 V^* \right)
\oplus
\left( S^2 U^* \otimes \Lambda^2 V^* \right)
\right)
\otimes
\left( V \otimes V^* \right)
\\
&\cong
\left( \Lambda^2 U^*
\otimes
\left( V \otimes V^* \otimes S^2 V^* \right)\right)
\oplus
\left(
S^2 U^*
\otimes
\left( V \otimes V^* \otimes \Lambda^2 V^* \right)\right).
\end{align*}
We now wish to consider the component of $\mathbf{R}^+$ which is a section of
$\Lambda^2 U^* \otimes \left( V \otimes V^* \otimes S^2 V^* \right)$. If
we completely symmetrise in the $V$ components, then the trace-free part
of the remaining object will be referred to as the positive part of the
Weyl tensor:
\[
W^+ \in \Gamma ( \mbox{Trace-free part of }
\Lambda^2 U^* \otimes \left( V^* \otimes S^3 V^* \right)).
\]

\begin{definition}
A $(p, q)$ paraconformal structure is right flat if
\begin{align*}
&\mathbf{T}^+ = 0, &p>2,\\
&W^+ = 0, &p=2.
\end{align*}
\end{definition}

In the case $p>2$, the vanishing of the torsion implies automatically
that $W^+ = 0$, whereas if $p=2$, the torsion $\mathbf{T}^+$
automatically vanishes, so the condition $W^+ = 0$ is non-trivial \cite{BE}.
A complex four dimensional spin-manifold with a metric is a
particular case of a paraconformal manifold with $p=q=2$ since,
due to the structure of the complexified rotation group,
the complexified tangent bundle as a product of spin-bundles.
In this case, $W^+$ may be identified with the self-dual part of
the Weyl tensor of the conformal structure \cite{AHS}.
In higher dimensions, with $p=2k$ and $q=2$,
special cases of right-flat paraconformal structures include
quaternion-K{\"a}hler and hyper-K{\"a}hler structures.

\subsection{Twistor Spaces}
\label{sec:twistor}
In the case of right-flat paraconformal structures,
there is an associated complex manifold $Z$ of dimension
$(p+1)(q-1)$ which defines the structure. This manifold
is constructed as follows.

Consider a $(p, q)$ paraconformal structure on a complex manifold $M$
as above, and assume we have a local basis
$\{ \boldsymbol{\epsilon}^a | a = 1, \dots, n \}$
for $T^* M$. The isomorphism~\eqref{paraisom} implies we may write
this as $\{ \epsilon^{AA^{\prime}} | A = 1, \dots , p ,
A^{\prime} = 1, \dots , q \}$. Given any $\pi_x \in V_x$,
we define the annihilator
\[
\pi^{\perp}_x := \{ \phi_x \in V^*_x | < \phi_x, \pi_x > = 0 \} \subset V^*_x.
\]
Let $\Lambda \subset \Omega(V)$ be the distribution
on the total space of the bundle $V$ generated by the $1$-forms
\begin{align}
\sigma^A &:=
\phi_{A^{\prime}} \epsilon^{AA^{\prime}},
\qquad \phi \in \Gamma (\pi^{\perp}),
\\
\sigma^{A^{\prime}} &:= d\pi^{A^{\prime}} +
\gamma^{A^{\prime}}{}_{B^{\prime}} \pi^{B^{\prime}},
\label{holforms1}
\end{align}
where $\boldsymbol{\gamma}^{A^{\prime}}{}_{B^{\prime}}$ denote the
components of the connection on $V$. Complex conjugation gives the
complex conjugate ideal ${\overline\Lambda}$.
The sub-bundle of $T (V)$ annihilated by $\Lambda$ and
${\overline\Lambda}$ is spanned by the distributions $T$ and ${\overline
T}$, where $T \subset T (V)$ is spanned by the vector fields
\begin{align}
v_{A} &:= \pi^{A^{\prime}} D_{AA^{\prime}},
\qquad
A = 1, \dots , p.
\label{Tvecs}
\end{align}

The distribution $\Lambda$ is closed
($d \Lambda \subset \Lambda^1 \wedge \Lambda$)
if and only if the
distribution $T$ is closed under
Lie Brackets $\left[ T, T \right] \subset T$.
It is straightforward to show from Eq.~\eqref{Tvecs} that,
given $\lambda, \chi \in \Gamma (U)$
\begin{equation}
\left[ \mathbf{v}_{\lambda \otimes \pi} , \mathbf{v}_{\chi \otimes \pi}
\right] \equiv
- \mathbf{T}(\lambda \otimes \pi, \chi \otimes \pi) +
\left( \mathbf{R}(\lambda \otimes \pi, \chi \otimes \pi) \pi
\right)^{A^{\prime}} \frac{\partial}{\partial \pi^{A^{\prime}}} \pmod T.
\label{rflat}
\end{equation}
Therefore, if we wish the space $T$ to be closed under Lie Bracket we
require
\begin{align}
< \phi, \mathbf{T}(\lambda \otimes \pi, \chi \otimes \pi)> &= 0, \qquad
\forall \phi \in \Gamma(\pi^{\perp}), \qquad \forall \lambda, \chi \in
\Gamma(U),
\\
\mathbf{R}(\lambda \otimes \pi, \chi \otimes \pi) \pi &= 0, \qquad
\forall \lambda, \chi \in \Gamma(U).
\label{rflat2}
\end{align}
It is straightforward to show that if we fix the connections
$\nabla$ on $U$ and $V$ so as to preserve the scale, as mentioned
in Section~\ref{sec:para}, then Eqs.~\eqref{rflat2} imply that the
paraconformal structure is right-flat. Therefore,
the distribution $T$ is integrable (equivalently
$\Lambda$ is closed under exterior differentiation)
if and only if the paraconformal structure is right-flat.

We wish to consider the projective version of this construction.
Treating the section $\pi$ as homogeneous coordinates on the
projective space $P_{q-1} \cong {\pv}_p$ for each $p \in M$,
we may introduce complex coordinates on the
region $U_1 = \{ \pi \in \mathbb{C}^q | \pi^1 \neq 0 \}$
\[
\lambda^i = \frac{\pi^i}{\pi^1}, \qquad i = 2, \dots q.
\]
The projections of the $1$-forms above are
\begin{align}
\sigma^A &:= \epsilon^{A1} + \lambda^2 \epsilon^{A2} + \dots +
\lambda^q \epsilon^{Aq}, \qquad A = 1, \dots , p,
\\
\sigma^i &:= d\lambda^i - A^i, \qquad i = 2, \dots, q,
\label{holforms}
\end{align}
where $\mathbf{A}$ is the projective version of the connection.
We again denote the distribution in {\pv} defined by these
$1$-forms by $\Lambda$. Similarly, a distribution,
again denoted $T \subset T (\pv)$, is spanned
by the projection of the vector fields~\eqref{Tvecs}
\begin{align}
\mathbf{v}_A &:= D_1 + \lambda^2 D_2 + \cdots + \lambda^q D_q,
\qquad A = 1, \dots , p.
\label{holvecs}
\end{align}
The distribution spanned by these vector fields is integrable if and
only if the paraconformal structure on $M$ is right-flat. The
integrability of this distribution implies we have a set of integrable
$p$-dimensional planes in {\pv}. Quotienting out {\pv} by this distribution%
\footnote{We are assuming
that there is nothing globally pathological about the fibration, and that
such a quotient operation is justified} %
therefore defines a quotient manifold $Z$
of dimension $(p+1)(q-1)$, which we denote by $Z$. We therefore have a
map $p : Z \to M$, where the image of a point in $Z$ is a
$p$-dimensional plane in {\pv} (in twistorial terminology,
an $\alpha$-plane). We can then define the distribution
$p^* \Lambda \subset \Lambda ( Z )$, which is involutive
on $Z$. Since the dimension of $p^* \Lambda$ equals the dimension of
$Z$, this ideal therefore determines an almost-complex structure on $Z$.
Moreover, since $p^* \Lambda$ is involutive, this almost-complex
structure is integrable. Therefore $Z$ is a complex manifold of dimension
$(p+1)(q-1)$.

We now wish to invert this process and construct the manifold $M$ from
a generic complex manifold $Z$. Given a point $p \in M$, its image in
the manifold $Z$ constructed above is a copy of $P_{q-1} \subset Z$,
corresponding to the fibre ${\pv}_p$. We therefore wish to reconstruct
the manifold $M$ as the parameter space of embedded $P_{q-1}$'s in $Z$.
In order to carry out this construction, we need to determine the
normal bundle, $N$, of such an embedded sub-manifold.

In the notation of Eq.~\eqref{holforms1}, the co-normal bundle, $N^*$,
is spanned by the forms $\{ \phi_A \boldsymbol{\epsilon}^{AA^{\prime}}
| \phi \in \pi^{\perp} \subset V^*, A = 1, \dots, p \}$. The co-normal
bundle is therefore isomorphic to $p$ copies of the bundle $\pi^{\perp}
\subset V^*$ which annihilates the element $\pi \in V$. Given $x \in
P_{q-1}$, $\pi_x$ is an element of the complex line in $C^{q}$
corresponding to the point $x$, i.e. an element of the
$L_x$, where $L$ denotes the tautological bundle $L := H^{-1}$.
We define the Universal Quotient
bundle, $Q$, so that the short sequence of vector bundles
\[
0 \rightarrow L \rightarrow \mathbb{C}^q \rightarrow Q \rightarrow 0
\]
is exact, where $\mathbb{C}^q$ denotes
the trivial rank $q$ vector bundle over
$P_{q-1}$. The bundle $\pi^{\perp}$
is therefore isomorphic to $Q^*$, the
dual of the quotient bundle.
{}From the fact that $Q \cong H \otimes TP_{q-1}$
\cite{GH}, we deduce that
\begin{equation}
N^* \cong \oplus_1^p \Omega^1 (1),
\label{conormalbundle}
\end{equation}
where, for a general manifold $X$,
$\Omega^r$ denotes the bundle of $r$-forms on
$X$, and in the particular case of
$X = P_{q-1}$, we define
\[
\Omega^r (k) := \Omega^r (P_n) \otimes H^k.
\]
The dual of Eq.~\eqref{conormalbundle} provides the normal bundle of
$P_{q-1} \subset Z$
\begin{equation}
N \cong \left( \oplus_1^p H^{-1} \right) \otimes T (P_{q-1}).
\label{normalbundle}
\end{equation}

The paraconformal manifold $M$
is reconstructed as the set of embedded
$P_{q-1}$'s in $Z$. Given that
we know the form of the normal bundle of
an embedded $P_{q-1}$ corresponding to a point $x \in M$,
the number of deformations of the projective
space follows from Kodaira's theorem:
If $H^1 (P_{q-1}, N) = 0$, then
the space of embedded $P_{q-1}$'s
is a complex analytic manifold $M$,
and the tangent space, $T_x M$, is isomorphic to
$H^0 (P_{q-1}, N)$. To calculate
these cohomology groups, we need some
results concerning vector bundles
over complex projective spaces \cite{OSS}.
Serre duality states that for a holomorphic
vector bundle $E$ over a (projective algebraic) complex $n$
manifold $X$ we have the isomorphism
\[
H^q (X, E) \cong \left( H^{n-q} (X, K_X \otimes E^*) \right)^*,
\]
where $K_X$ denotes the canonical bundle of $X$.
On such a manifold we also have the identification
\[
(\Omega^r)^* \cong (\Omega^n)^* \otimes \Omega^{n-r}.
\]
For a complex projective space
\[
K_{P_n} \cong H^{-(n+1)},
\]
so in this case we have
\begin{equation}
H^q (P_n, \Omega^p (k)) \cong
\left( H^{n-q} (P_n, \Omega^{n-p} (-k) \right)^*.
\label{pncohom}
\end{equation}
Results of Bott \cite{Bo} then tell us that
\begin{equation}
\mathrm{dim}_{\mathbb{C}} H^q (P_n, \Omega^p (k)) =
\begin{cases}
\binom{n+k-p}{k} \binom{k-1}{p}&
q=0, 0 \le p \le n, k > p,
\\
1& k=0, 0 \le p = q \le n,
\\
\binom{-k+p}{-k} \binom{-k-1}{n-p}&
q=n, 0 \le p \le n, k < p-n,
\\
0& \mbox{otherwise}.
\end{cases}
\label{bott}
\end{equation}

Applying these results, we first
show that $H^1 (P_{q-1}, N) = 0$.
{}From Eqs.~\eqref{conormalbundle}
and \eqref{pncohom}, we find that
\begin{align*}
H^1 (P_{q-1}, N) &\cong
\left( H^{q-2} (P_{q-1}, K \otimes N^*) \right)^*
\\
&\cong
\left( H^{q-2} (P_{q-1}, H^{-q} \otimes \oplus_1^p \Omega^1 (1))
\right)^*
\\
&\cong
\oplus_1^p \left( H^{q-2} (P_{q-1}, \Omega^1 (1-q)) \right)^*
\\
&\cong
\oplus_1^p H^1 (P_{q-1}, \Omega^{q-2} (q-1))
\\
&\cong 0,
\end{align*}
where the last equality follows
from Eq.~\eqref{bott}. Therefore $T_x M$ is
isomorphic to $H^0 (P_{q-1}, N)$ where
\begin{align*}
H^0 (P_{q-1}, N) &\cong
\oplus_1^p H^0 (P_{q-1}, \Omega^{q-2} (q-1))
\\
&\cong \mathbb{C}^{pq},
\end{align*}
by a similar argument to that given above.
Therefore, given an embedded $P_{q-1}$ in a complex manifold $Z$ of
complex dimension $(p+1)(q-1)$, with normal bundle as in
Eq.\eqref{normalbundle}, there
will exist a $n=pq$ parameter family of
such spaces. In the usual fashion, the
integrability of the complex structure on $Z$ then implies that $M$
carries a right-flat paraconformal structure.

\section{Integrable Systems Interpretation}
\label{sec:integrable}

The integrability of the
distribution $T$ defined by the vector
fields~\eqref{holvecs} implies the existence of
functions $C_{AB}{}^C$ with the property that
\begin{equation}
\left[ \mathbf{v}_A, \mathbf{v}_B \right] =
\sum_{C=1}^p C_{AB}{}^C \mathbf{v}_C, \qquad
A, B = 1, \dots , p,
\label{laxequations}
\end{equation}
where, we recall, the vector fields $\mathbf{v}_A$ are defined by
\[
\mathbf{v}_A = D_1 + \lambda^2 D_2 + \cdots + \lambda^q D_q,
\]
with
\[
D_i = \mathbf{e}_i - A_i{}^j
\frac{\partial}{\partial {\lambda}^j},
\]
with the $A_i{}^j$ quadratic
polynomials in the spectral parameters
$\lambda^i$. As such, the $p$ vector fields $\mathbf{v}_A$
are sections of the tangent bundle of
$P(V)$, and correspond to
differential operators which depend on a set
of $(q-1)$ spectral parameters
$(\lambda^2, \dots, \lambda^p)$. More properly, these
parameters correspond to a
section of the line bundle $H$ over $P_{q-1}$, so
our \lq\lq spectral parameter\rq\rq
now lives in $P_{q-1}$, unlike the
usual case where we have a single spectral parameter in
$P_1$. A power counting argument implies that the
functions $C_{AB}{}^C$ are
quadratic polynomials in the complex
coordinates $\lambda^i$, corresponding to
sections of the bundle $H^2$.

As in the description of
anti-self-dual conformal structures in
four dimensions, the differential
operators $\mathbf{v}_A$ contain
partial derivatives with
respect to these spectral parameters,
corresponding to the fact that
the complex manifold $Z$ generally does
not fibre holomorphically over $P_{q-1}$.

The integrability of the
distribution $T$ is equivalent to the
fact that the differential
ideal $\Lambda$ is involutive. Integrability
of $T$ implies the integrability
of a distribution of $p$-dimensional
planes in $P(V)$, and the existence
of $(p+1)(q-1)$ functions $f^{\alpha}$ such that the
planes are level sets of these
functions. Equivalently the differential
ideal $\Lambda$ is generated by
the differentials $\{ df^{\alpha} \}$. If we
then quotient out by the
$p$-dimensional distribution to construct the
manifold $Z$, then the functions
$f^{\alpha}$ descend to holomorphic functions on
the manifold $Z$, and $\{ df^{\alpha} \}$
generate $\Lambda^{(1, 0)} (Z)$.

In terms of the paraconformal manifold $M$, the
$p (p-1) /2$ equations~\eqref{laxequations}
are the integrability condition for over-determined set of
equations for a function $f(x: \mathbf{\lambda})$:
\begin{equation}
\mathbf{v}_A f = 0, \qquad A = 1, \dots p.
\label{associatedlinear}
\end{equation}
When Eqs.~\eqref{laxequations} are satisfied, there exist $(p+1)(q-1)$
linearly independent solutions of these equations $\{ f_{\alpha} \}$.
The sub-space $\{ f_{\alpha} = \mathrm{constant} \} \subset P(V)$ are
then the $\alpha$ planes of our right-flat paraconformal structure.
The functions $\{ f_{\alpha} \}$ then descend to holomorphic functions
on the quotient manifold $Z$.

In integrable systems terminology,
Eq.~\eqref{associatedlinear} is the
associated linear problem for
the right-flat paraconformal structure.
The compatibility condition
Eq.~\eqref{laxequations} then ensures the
integrability of the system.
There are several non-standard elements of
this construction, however.
Firstly, the analogue of the
spectral parameter of standard integrable
systems theory in these
equations is the set of affine coordinates
$\{ \lambda^i \}$ on the
$(q-1)$-dimensional complex projective space
$P_{q-1}$.

Secondly, as opposed to the usual
\lq\lq Lax Pair\rq\rq\ formulation of
integrable systems, we are here forced to consider a
\lq\lq Lax $p$-tuple\rq\rq\ of operators i.e. the vector fields
$\mathbf{v}_A$, which must
define an integrable distribution for the
complex structure on the manifold $Z$ to be integrable.

As in the simpler case of anti-self-dual conformal
structures in dimension $4$
(and similarly $3$-dimensional Einstein-Weyl
structures), the differential operators we consider
generally contain derivatives
in the spectral projective space,
corresponding to the fact that
the complex manifold $Z$ generally does not
fibre over the complex projective space $P_{q-1}$.

\subsection{Relations with Ward's Systems}

Equations~\eqref{laxequations}
are, in some sense, an analogue
of a construction due to Ward
for gauge fields \cite{W1}. Ward
considered principal
$G$-bundles with a connection
$\mathbf{A} \in \Gamma
(\Lambda^1 \otimes \mathfrak{g})$.
We consider a field $\psi$
in a representation of $G$, and
consider the over-determined
set of linear equations
\begin{equation}
D_{\mathbf{V}_\alpha} \psi = 0,
\qquad
\alpha = 1, \dots, p
\label{ward1}
\end{equation}
where $D \psi$ denotes the
covariant derivative of the field $\psi$
with respect to the connection $\mathbf{A}$, and the
$\mathbf{V}_{\alpha}$ are vector
fields. Moreover, the vector fields
$\mathbf{V}_{\alpha}$ are taken to depend on a set
of complex parameters
$\{ \lambda^A | A = 1, \dots, q \}$, being
a homogeneous polynomial of
degree $N$ in these parameters.
(The vector fields may therefore
be identified with a section of
$TM \otimes H^N$, where $H$
denotes the Hopf bundle over the
complex projective space $P_{q-1}$.)
For fixed $\lambda^A$, Eqs.~\eqref{ward1} are
actually $p \dim \mathfrak{g}$
differential equations for
$\dim \mathfrak{g}$ unknowns,
and so are over-determined if
$p > 1$. Since
the system is over-determined,
the existence of a maximal family
of solutions places a set of algebraic
constraints on the curvature
$\mathbf{F}$ of the connection
\begin{equation}
\mathbf{F} (\mathbf{V}_\alpha, \mathbf{V}_\beta) = 0,
\qquad
\alpha, \beta = 1, \dots, p.
\label{ward2}
\end{equation}
In the cases where the set of
polynomial vector fields $\mathbf{V}_{\alpha}$
are suitably non-degenerate,
the equations \eqref{ward2}
can be completely solved by
twistorial techniques \cite{W1}.

The connection with paraconformal
structures arises if we consider
the case of linear polynomials 
corresponding to $N=1$. In this case, the
non-degeneracy condition mentioned
above is analogous to the defining
isomorphism~\eqref{paraisom}.
If we assume the underlying manifold
of the theory is $\mathbb{R}^{pq}$, with
coordinates $\{ x^a : a = 1, \dots , pq \}$,
and that the connection is constant
(i.e. independent of the $x^a$),
then the integrability
conditions above become a set of
algebraic equations on the connection
\[
\left[ \mathbf{A} (\mathbf{V}_{\alpha}),
\mathbf{A} (\mathbf{V}_{\alpha}) \right] = 0,
\qquad
\alpha, \beta = 1, \dots, p.
\]
If we now take the connection $\mathbf{A}$ to have values in the
tangent bundle of an auxiliary manifold ${\overline M}$, then we may
write
\begin{equation}
\mathbf{A} (\mathbf{V}_{\alpha}) =
\sum_{a=1}^n \sum_{A=1}^q \sigma_{\alpha A}^a \lambda^A
\mathbf{e}_a
\end{equation}
where $\{ \mathbf{e}_a | a = 1, \dots, n \}$ denotes
a basis of vector fields on the manifold ${\overline M}$.
The integrability conditions~\eqref{ward2} then
reduce to a set of relations on the commutators of the
vector fields $\{ \mathbf{e}_a \}$ on ${\overline M}$
\begin{equation}
\sum_{A, B = 0}^q \sum_{a, b = 1}^n
\lambda^A \lambda^B
\sigma_{\alpha A}^a \sigma_{\beta B}^b
\left[ \mathbf{e}_a, \mathbf{e}_b \right]  = 0,
\label{ward3}
\end{equation}
where $\left[ \ , \ \right]$ denotes
the Lie bracket of vector fields. 
Imposing Eqs.~\eqref{ward3} for all values of the parameters 
$\lambda^A$, we recover Eqs.~\eqref{laxequations}
with $C_{AB}{}^C = 0$ in the case when
the derivatives with respect to the spectral parameter are not
present. 

If we allow derivatives with respect to the spectral
parameter, then, in the terminology of Park \cite{P},
our equations for general
right-flat paraconformal structures may therefore be considered
as a $P_{q-1}$-extension of Ward's equations for a constant 
connection on flat space with values in the tangent bundle 
of an auxiliary $n$-manifold ${\overline M}$. 

A special case of these equations without derivatives 
with respect to the spectral parameters 
is Joyce's interpretation of the equations
for hyper-complex conformal structures in
four dimensions \cite{J}, which in turn
is a generalisation of the description of
anti-self-dual Ricci-flat structures due
to Mason and Newman \cite{MN}. 

In terms of Ward's classification of systems in
dimensions up to $11$, paraconformal structures of
type $p=k, q=2$ correspond to Ward's systems $A_k$,
$p=2, q=m+1$ correspond to his $C_m$,
and $p=q=3$ correspond to his system $D$.
The geometrical analogue of Ward's systems
with higher order homogeneous polynomials 
correspond to twistor spaces $Z$
containing embedded $P_{q-1}$'s with
more complicated normal bundle, 
sections of which can be identified with
a collection of sections of the 
bundle $H^n$. Unfortunately, there does
not seem to be any simple geometrical 
interpretation of these systems
in general.

\section{Remarks and Conclusions}

If we wish to take seriously the idea that at the heart of classical
integrable systems is a connection with complex geometry, then implicit
in the formulation of paraconformal structures given in
Eq.~\eqref{laxequations} are several generalisations of standard notions
of integrability.

Firstly, the analogue of the spectral parameter of standard integrable
systems theory in these equations is the set of affine coordinates
$\{ \lambda^i \}$ on the $(q-1)$-dimensional complex projective space
$P_{q-1}$. In other words, the spectral parameter lives in a general
complex projective space.

Secondly, as opposed to the usual \lq\lq Lax Pair\rq\rq\ formulation of
integrable systems, we are here forced to consider a
\lq\lq Lax $p$-tuple\rq\rq\ of operators i.e. the vector fields
$\mathbf{v}_A$, which must define an integrable distribution for the
complex structure on the manifold $Z$ to be integrable.

Finally, as in the simpler case of anti-self-dual conformal
structures in dimension $4$ (and similarly $3$-dimensional Einstein-Weyl
structures), the differential operators we consider
generally contain derivatives in the spectral projective space,
corresponding to the fact that the complex manifold $Z$ generally does not
fibre over the complex projective space $P_{q-1}$.

The only case in which we recover a standard Lax-pair construction with
spectral parameter in $P_{1}$ is the case $(p, q) = (2, 2)$, when the
twistor space fibres over $P_{q}$. Geometrically, this corresponds to
the description of (complexified) hyper-complex
structures in four dimensions.

The second observation above is consistent with the complex-manifold approach
to hyper-complex and quaternionic-k{\"a}hler manifolds of real dimension
$4k$, where the points of the manifold correspond to rational curves
($q=2$) with normal bundle $\oplus_1^{2k} H$ in a complex manifold $Z$ of
dimension $2k+1$. One could further generalise this picture by considering
more general embedded complex sub-manifolds than $P_{q-1}$, with more complicated normal
bundles. The geometrical structures induced on the space of such sub-manifolds is, however,
rather unclear. Even if we restrict ourselves to embedded rational curves,
Grothendieck's Theorem implies that the
most general normal bundle is of the form
$N \cong \oplus_{i=1}^n H^{m_i}$ for integers $m_i$, but
the geometrical interpretation of
the induced structure on the space of rational curves
for general $m_i$ is far from apparent. The only case which is known to have a sensible
geometrical description is the description of $3$-dimensional Einstein-Weyl structures, where
we have a complex $2$-manifold $Z$, and a family of rational curves with normal bundle
$N \cong H^2$ \cite{H}.

Finally, we should note that we have
considered only complex paraconformal structures.
If we consider an analytic real paraconformal manifold,
where the complexified tangent bundle splits as a tensor
product, then we may
complexify the manifold and use the complex construction of the twistor
space given above. However, there does not seem to be any straightforward
definition of the twistor space in the case of non-analytic real
paraconformal manifolds. The hope would be that analyticity
follows from existence of a right-flat paraconformal structure, in the
same way that in $4$-dimensions the existence of an anti-self-dual
conformal structure implies the existence of a real analytic
structure \cite{AHS}. {}From the twistorial point
of view, we require that the complex manifold $Z$ admit
a real structure (i.e. an anti-holomorphic involution), $\sigma$,
and that there be a $n$-parameter family of real
embedded $P_{q-1}$'s which are invariant under this map.
These invariant $P_{q-1}$'s then correspond to points of
the manifold $M$. Since the manifold $M$ is then a real sub-manifold
of a complex-analytic manifold, it then necessarily admits a
real-analytic structure. The existence (or not) of fixed
points of $\sigma$ then allows us to attribute a signature
to the induced paraconformal structure on $M$, with the
fixed point set generically defining a real projective
space, which determines the set of null planes at a given
point in $M$. A real structure on $Z$ with no
fixed points would define the analogue of a Riemannian
structure of Theorem~\ref{thm:ahs}.

\begin{ack}
The author is grateful
to I.A.B. Strachan for
useful conversations relating
to anti-self-dual conformal structures,
and for pointing out an error
in a previous version of this paper.
This work was funded by the EPSRC, and
the University of Hull.
\end{ack}

\end{document}